\documentclass[fleqn]{2023SCGE}
\usepackage{indentfirst}
\usepackage{footmisc}
\usepackage{hyperref}
\usepackage[toc]{multitoc}

\begin{document}

\textcolor[rgb]{0.88,0.77,0}{Table}
\ensubject{subject}

\ArticleType{Article}
\SpecialTopic{SPECIAL TOPIC: }
\Year{2026}
\Month{}
\Vol{00}
\No{0}
\DOI{??}
\ArtNo{000000}
\ReceiveDate{April 29, 2026}
\AcceptDate{}

\title{Electronic and topological properties of Ce-based honeycomb ferromagnet Ce$_2$Zn$_6$Ge$_3$}{Electronic and topological properties of Ce-based honeycomb ferromagnet Ce$_2$Zn$_6$Ge$_3$}

\author[1]{Yanen Huang\footnotemark[2]}{}

\author[1]{Zihan Yang\footnotemark[2]}{}

\author[1]{Jiawen Zhang}{}

\author[1]{Yuwei Zhou}{}

\author[1]{Lubin Wang}{}

\author[2,3]{Gang Li}{}

\author[1]{\\Michael Smidman}{}

\author[1]{Chao Cao}{}

\author[1]{Yu Liu}{{liuyuccm@zju.edu.cn}}

\author[1,4,5,6]{Huiqiu Yuan}{{hqyuan@zju.edu.cn}}

\AuthorMark{Y. E. Huang}

\AuthorCitation{Y.E. Huang, Z. H. Yang, et al}

\address[1]{New Cornerstone Science Laboratory, Center for Correlated Matter and School of Physics, Zhejiang University, Hangzhou 310058, China}
\address[2]{Beijing National Laboratory for Condensed Matter Physics, Institute of Physics, Chinese Academy of Sciences, Beijing 100190, China}
\address[3]{School of Physical Sciences, University of Chinese Academy of Sciences, Beijing 100190, China}
\address[4]{Institute of Fundamental and Transdisciplinary Research, Zhejiang University, Hangzhou 310058, China.}
\address[5]{State Key Laboratory of Silicon and Advanced Semiconductor Materials, Zhejiang University, Hangzhou 310058, China.}
\address[6]{Collaborative Innovation Center of Advanced Microstructures, Nanjing 210093, China.}

\AuthorMark{Y.E. Huang}

\AuthorCitation{Y. E. Huang, Z. H. Yang, et al}


\abstract{Ce$_2$Zn$_6$Ge$_3$ is a rare example of the Ce-based honeycomb ferromagnet. Here, we report its Fermi surface and topological properties by quantum oscillations via the magnetoresistance and tunnel diode oscillator (TDO) based measurements, in combination with the density functional theory (DFT) calculations. Four fundamental frequencies are observed in the quantum oscillations, and their angle dependence is more compatible with the DFT calculations assuming that the 4$f$-electrons are localized, suggesting a localized nature of ferromagnetism in Ce$_2$Zn$_6$Ge$_3$. Furthermore, the observations of negative longitudinal magnetoresistance and non-zero Berry phase, as well as the existence of two pairs of Weyl points near the Fermi level as revealed from the calculated electronic structures, provide strong evidence for nontrivial topology in Ce$_2$Zn$_6$Ge$_3$. These findings suggest that Ce$_2$Zn$_6$Ge$_3$ could provide a unique platform to study magnetism, topology, quantum criticality and their interplay.}



\maketitle


\begin{multicols}{2}
\section{Introduction}\label{section1}

The honeycomb lattice, consisting of two interpenetrating triangular sublattices, is characterized by a $D_{6h}$ point group symmetry that fundamentally dictates its physical properties. In the non-interacting limit, this geometry hosts massless Dirac fermions at the $K$ and $K'$ points of the hexagonal Brillouin zone \cite{honeycomb_topo_RevModPhys.81.109}. Furthermore, Haldane demonstrated that quantum anomalous Hall effect without Landau levels can be realized in the honeycomb lattice \cite{honeycomb_Haldane_PRL1988}. Subsequently, it 
\Authorfootnote

\noindent  
was found that the inclusion of intrinsic spin-orbit coupling in this lattice can induce the quantum spin Hall effect \cite{Topo_RMP2010,Topo_RMP2011}. Beyond these nontrivial electronic properties, the honeycomb lattice also serves as a platform for exploring exotic magnetic states. For example, the isotropic nearest-neighbor exchange interactions in the honeycomb lattice may establish a long-range magnetic order \cite{long_range_order_PhysRevLett.110.097204,long_range_order_PhysRevLett.108.127203,long_range_order_PhysRevB.107.064409}, while the competing further-neighbor exchange or anisotropic exchange interactions can induce strong magnetic frustrations, potentially suppressing conventional order and favoring a quantum spin liquid state \cite{QSL_Tokiwa2014,QSL_PhysRevX.12.031039,QSL_Meng2010}. Rare-earth-based honeycomb magnets are particularly noteworthy, as the strong electronic correlations and

\begin{figure}[H]
	\includegraphics[width=0.9\columnwidth]{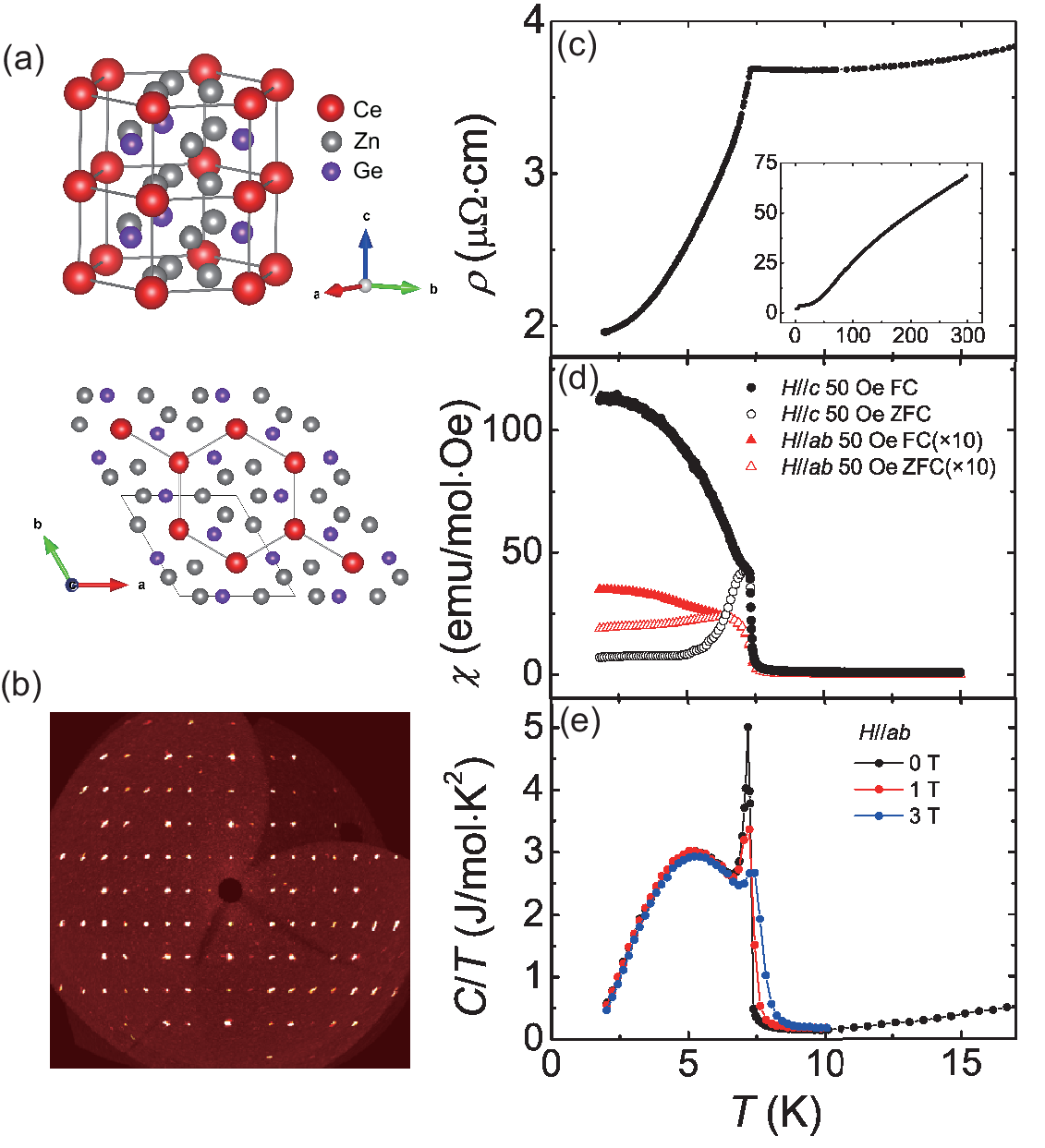} 
	\centering
	\caption{
		Characterization of the basic physical properties of Ce$_2$Zn$_6$Ge$_3$. (a) The crystal structure of Ce$_2$Zn$_6$Ge$_3$. The Ce ions in the $ab$-plane form a honeycomb lattice. (b) The single-crystal X-ray diffraction (XRD) result of Ce$_2$Zn$_6$Ge$_3$. (c) Temperature dependence of the electrical resistivity at low temperatures; the inset shows the resistivity from 2~K to 300~K. (d) The magnetic susceptibility $\chi (T)$ with a magnetic field of 50 Oe applied along the $c$-axis (black circle) and perpendicular to the $c$-axis (red triangle). The open and filled symbols represent the  zero-field cooling (ZFC) and field-cooling (FC) processes, respectively. (e) The low-temperature specific heat plotted as $C(T)/T$ with different fields applied along the $ab$-plane.
	}
	\label{Fig_basic_properties}
\end{figure}

\noindent
magnetism arising from $f$ electrons may intertwine with nontrivial band topology.

Ce$_2$Zn$_6$Ge$_3$ is a rare example of a 4$f$-electron-based honeycomb ferromagnet. Here the Ce atoms form a honeycomb plane, which may host nontrivial topological properties. Besides, this compound undergoes a ferromagnetic (FM) transition at $T_{\rm C} \approx 7$~K with a much reduced saturated magnetic moment, similar to CeRh$_6$Ge$_4$ where a pressure-induced ferromagnetic quantum critical point was recently observed near 0.8 GPa \cite{CeRh6Ge4_Shen2020}. It has been theoretically proposed and experimentally demonstrated in CeRh$_6$Ge$_4$ \cite{CeRh6Ge4_Shen2020,FMQCPtheory_PhysRevLett.126.216406,FS_compare_scibullet2021,FMQCPtheory_PRL,FMQCPtheory_SciChina} that the localized ferromagnetism with a pronounced magnetic anisotropy is crucial for the development of a ferromagnetic quantum phase transition. Therefore, it is important to further characterize the electronic and magnetic properties of Ce$_2$Zn$_6$Ge$_3$ in order to explore the possible ferromagnetic quantum criticality in this compound. Moreover, despite the honeycomb arrangement of Ce atoms and the presence of strong spin-orbit coupling, the topological properties of Ce$_2$Zn$_6$Ge$_3$ remain largely unexplored. Since the Fermi surface serves as a key probe of both the itinerancy of Ce-4$f$ electrons and the

\begin{figure}[H]
	\centering
	\includegraphics[width=1\columnwidth]{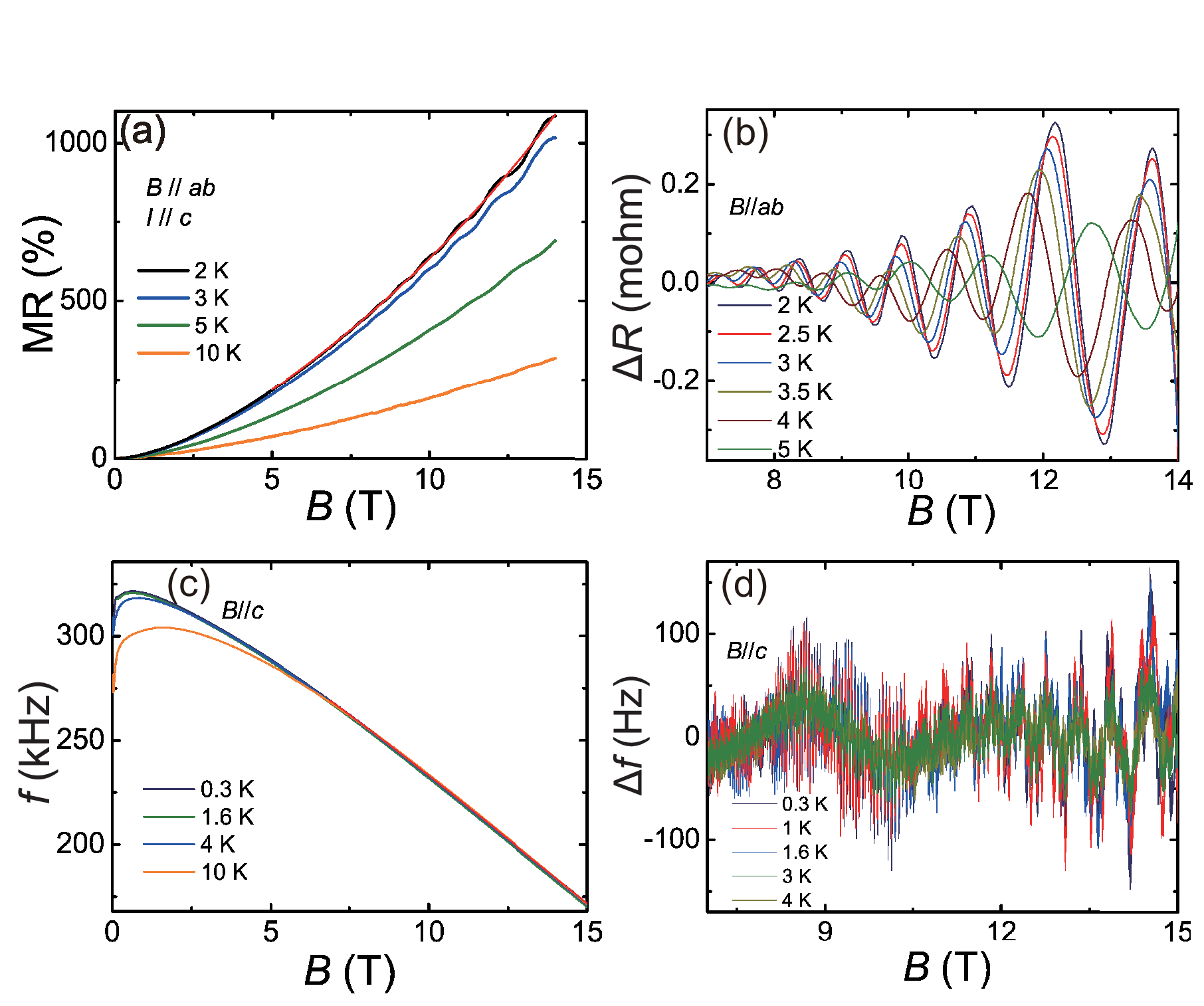} 
	\protect\caption{
		Quantum oscillations of	Ce$_2$Zn$_6$Ge$_3$. (a) The MR of Ce$_2$Zn$_6$Ge$_3$ with the field along the $ab$-plane, and the red solid line is the polynomial fit to the data at 2 K. (b) Quantum oscillations of MR after subtracting a polynomial background. (c) The TDO results with the field along the $c$-axis, and the red solid line is the polynomial fit to the data at 0.3 K. (d) Quantum oscillations of TDO data after subtracting a polynomial background. 
	}
	\label{Fig_QO_raw_data}
\end{figure}

\noindent
 nature of the topological states, a detailed investigation of its electronic structure is highly desirable.

In this work, we investigate the Fermi surface and topological properties of Ce$_2$Zn$_6$Ge$_3$ through measurements of the magnetoresistance (MR) and the TDO-based quantum oscillations with a magnetic field up to 30 T and a temperature down to 60 mK, together with the DFT calculations of band structure. The experimentally derived quantum oscillation frequencies are compatible with the theoretical scenario assuming the 4$f$ electrons are localized, suggesting localized ferromagnetism in Ce$_2$Zn$_6$Ge$_3$. Furthermore, observations of a negative longitudinal MR, a nontrivial Berry phase extracted from the quantum oscillations, as well as the observation of two pairs of Weyl points near the Fermi level in the calculated band structure provide strong evidence for the existence of a nontrivial topological state in Ce$_2$Zn$_6$Ge$_3$. These findings establish Ce$_2$Zn$_6$Ge$_3$ as a promising platform for investigating the interplay among topology, magnetism, and quantum criticality.

\section{Experimental methods}\label{sec:2}

Single crystals of Ce$_2$Zn$_6$Ge$_3$ were grown using the excess Zn-flux method. The cerium ingot (Alfa, 99.8\%), zinc granules (PrMat, 99.999\%), and germanium granules (PrMat, 99.9999\%) were loaded into an alumina crucible with a molar ratio of Ce:Zn:Ge = 2:200:3. The crucible was then sealed

\begin{figure}[H]	\includegraphics[width=1\columnwidth]{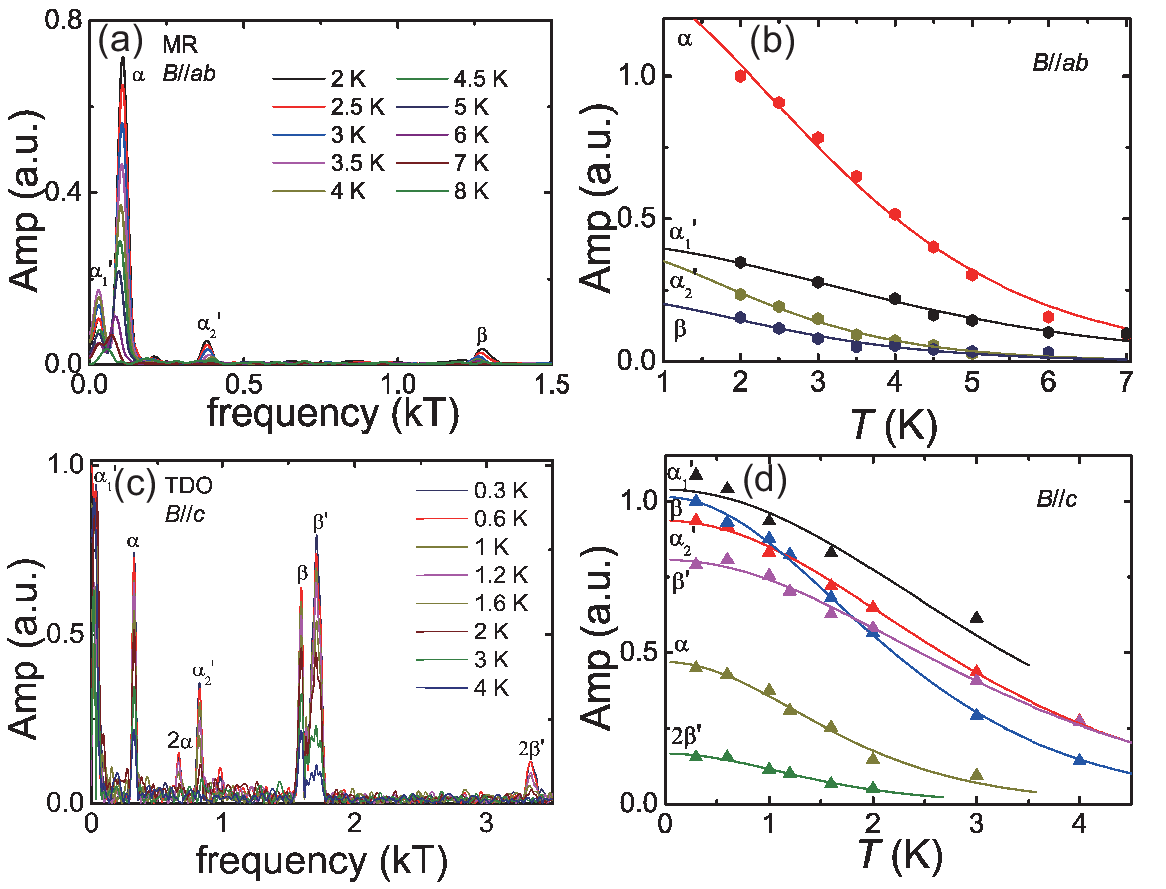} 
	\protect\caption{
		(a) and (c): The FFT analyses of the quantum oscillations shown in Fig. \ref{Fig_QO_raw_data}; (b) and (d): Temperature dependence of their oscillation amplitudes. (a) and (b) are based on the MR measurements; (c) and (d) are derived from the TDO data. The solid lines in (b) and (d) represent the fits by using the Lifshitz-Kosevich formula, as described in the main text.  
	}
	\label{Fig_QO_and_FFT}
\end{figure} 

\begin{table}[H]
	\begin{threeparttable}
		\caption{Experimentally determined effective masses $m^*$ for magnetic fields within the $ab$-plane (MR measurements) and parallel to the $c$-axis (TDO measurements).}
		\label{tab1}
		\doublerulesep 0.1pt \tabcolsep 5.5pt
		\begin{tabular}{c c c c c}
			\toprule
			band & frequency  & $m^*$  & frequency  & $m^*$  \\
			\ & (kT, $B$$\parallel$$ab$) & ($m_{\rm{e}}$, $B$$\parallel$$ab$) & (kT, $B$$\parallel$$c$) & ($m_{\rm{e}}$, $B$$\parallel$$c$) \\
			\hline\\[-2ex]
			$\alpha_1 '$ & 0.03 & 0.37(2) & 0.03 & 0.38(5)\\
			$\alpha$ & 0.11 & 0.47(1) & 0.32 & 0.42(1)\\
			$\alpha_2 '$ & 0.32 & 0.65(3) & 0.81 & 0.55(2)\\
			$\beta$ & 1.30 & 0.59(6) & 1.59 & 0.52(1)\\
			$\beta '$ & \ & \ & 1.71 & 0.70(1)\\
			2$\beta '$ & \ & \ & 3.39 & 0.55(2)\\
			\bottomrule
		\end{tabular}
	\end{threeparttable}
\end{table}

\noindent
in an evacuated quartz ampoule containing approximately 0.05 atm of high-purity argon. The ampoule was heated to 900~$^\circ$C, held at this temperature for 12 h, and subsequently cooled slowly to 500~$^\circ$C. Shiny single crystals were obtained by centrifuging the excess flux. The phase purity and chemical composition were confirmed by single-crystal X-ray diffraction (XRD)[Fig.~\ref{Fig_basic_properties}(b)] and energy-dispersive X-ray spectroscopy (EDS) performed using a field-emission scanning electron microscope (SEM).

 Resistivity and heat capacity measurements were conducted using a Quantum Design Physical Property Measurement System (PPMS) under magnetic fields up to 14~T. TDO measurements were performed in an Oxford HelioxVL $^3$He refrigerator in fields up to 15~T. Experiments at higher magnetic fields (up to 30~T) were carried out in a top-loading dilution refrigerator using the superconducting magnet at the Synergetic Extreme Condition User Facility (SECUF) in Beijing. For the TDO measurements, a homemade coil was wrapped firmly around the sample~\cite{tdo_huang2022}, operating at resonant frequencies of approximately 25~MHz for measurements with a magnetic field up to 15 T and 28~MHz for high-field measurements, respectively.

First-principles calculations were performed within the framework of the generalized gradient approximation (GGA) with the Perdew-Burke-Ernzerhof (PBE) functional~\cite{GGA_PRL1996}, as implemented in the VASP package~\cite{VASP_PRB1996}. Maximally projected Wannier functions were constructed using the WANNIER90 code~\cite{Wannier_CPC2008} and symmetrized via WANNSYMM~\cite{Wannsymm_CPC2022}. Surface states were calculated with WANNIERTOOLS~\cite{Wanniertools_CPC2018}, and Fermi surfaces were visualized using XCrySDen~\cite{XCrySden_CMS2003,Xcrysden_JMGM1999}.

\section{Result and discussion}\label{sec:3}

\subsection{Basic properties and quantum oscillations}

The unit cell of Ce$_2$Zn$_6$Ge$_3$ is illustrated in Fig.~\ref{Fig_basic_properties}(a). The compound crystallizes in a hexagonal structure belonging to the $P\bar{6}2m$ space group, where the Ce ions form a honeycomb lattice in the $ab$-plane. Figure~\ref{Fig_basic_properties}(c) shows the temperature dependence of the resistivity $\rho(T)$ at low temperatures. The resistivity $\rho$(T) shows a distinct anomaly at $T_{\rm C} \approx 7$~K, which corresponds to the FM transition. A large residual resistivity ratio [RRR = $\rho(300~\rm{K})/\rho(2~\rm{K})$] of about 35 is observed, suggesting significantly improved crystal quality compared to earlier studies~\cite{AGrytsiv_2003}. Figure~\ref{Fig_basic_properties}(d) presents the magnetic susceptibility $\chi(T)$ measured with a magnetic field of 50~Oe applied along the $ab$-plane and the $c$-axis, respectively. The susceptibility confirms a FM transition at 7~K, with a clear bifurcation between the FC and ZFC curves for both orientations; the $c$-axis is identified as the easy magnetization axis, in contrast to CeRh$_6$Ge$_4$ where the easy axis lies within the $ab$-plane. As shown in Fig.~\ref{Fig_basic_properties}(e), a pronounced jump in the specific heat is evident at $T_{\rm C}$. The crystal electric field (CEF) in the hexagonal environment lifts the ground-state degeneracy~\cite{CEF_1977}, giving rise to a Schottky anomaly at lower temperatures and leading to a broad hump in the specific heat below $T_{\rm C}$, which displays a slight shift toward higher temperatures under fields within the $ab$-plane.

To investigate its Fermi surface, magnetoresistance and TDO measurements were systematically performed. Figures~\ref{Fig_QO_raw_data}(a) and \ref{Fig_QO_raw_data}(b) display the MR results and the oscillatory component after background subtraction, respectively, show-

\begin{figure}[H]
	\includegraphics[width=1\columnwidth]{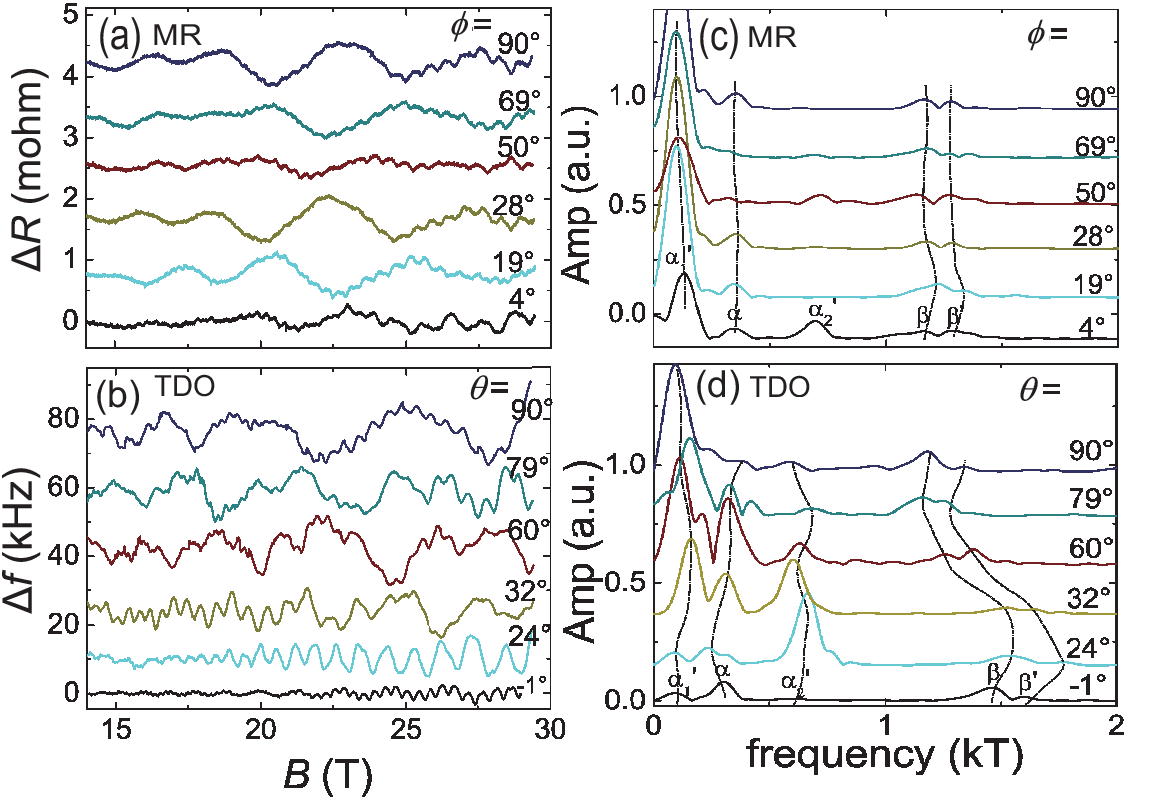} 
	\protect\caption{
		High-field quantum oscillations at 60 mK, with a background subtracted, are displayed for (a) different angles $\phi$ within the $ab$-plane, with $\phi = 0$ corresponding to the $a$-axis, and (b) different angles $\theta$ relative to the $c$-axis. (c) and (d) show the FFT results for (a) and (b), respectively. Two slices, carefully co-sliced from a single, large, high-quality batch crystal, were utilized for separate sets of measurements.
	}
	\label{Fig_high_field_QO}
	
\end{figure}

\noindent
ing clear Shubnikov-de Haas (SdH) oscillations at low temperatures. Here, the MR ratio is defined as MR = $[R(B)-R(0)]/{R(0)}$, where $R(B)$ and $R(0)$ are the resistance measured with and without magnetic 
field $B$, respectively. A large, non-saturating MR is observed at low temperatures, similar to the behavior reported in some topological semimetals with compensated electron and hole carriers~\cite{compensated_nature_Ali2014,compensated_semimetal_PRL2016,compensated_semimetal_PRB2017,compensated_Du2016,compensated_CundongLi_CPB_97307}. Due to the needle-like morphology of the single crystals, TDO measurements are particularly effective for fields oriented along the $c$-axis [Fig. \ref{Fig_QO_raw_data}(c)]. This contactless technique is more sensitive for detecting the high-frequency oscillations, as shown in Fig. \ref{Fig_QO_raw_data}(d). 

\begin{figure*}[t]
	\includegraphics[width=1.65\columnwidth]{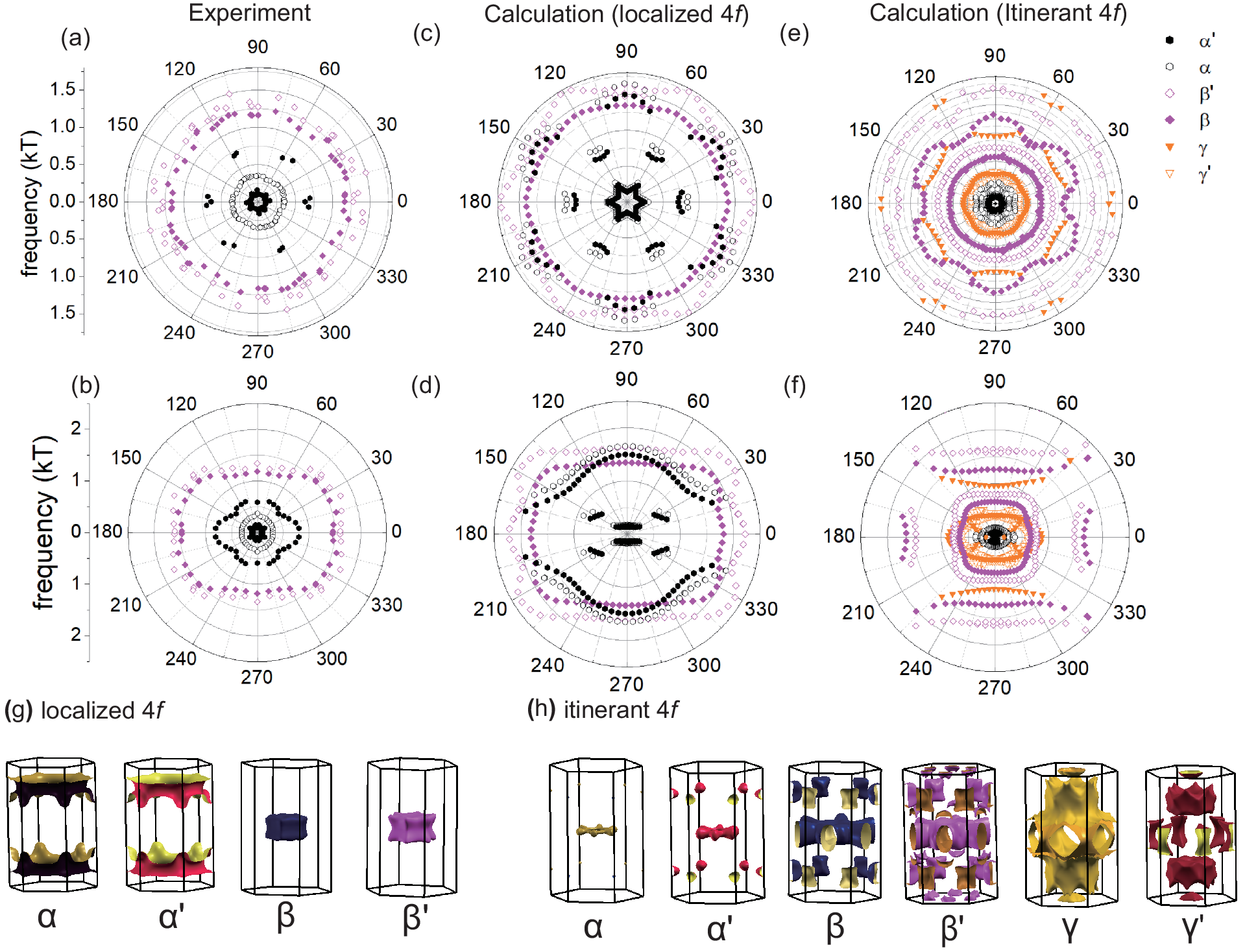} 
	\centering\protect\caption{
		Comparison between the experimentally derived quantum oscillation frequencies (a, b) and those of the DFT calculations assuming the Ce-4$f$ electrons are either localized (c, d) or itinerant (e, f).  The frequencies are plotted as a function of the in-plane field angles ($\phi$) (a, c, e) and the out-of-plane field angles ($\theta$) (b, d, f), respectively. The frequencies derived from different Fermi-surface sheets are labeled by different symbols. Note that multiple frequencies may come from the same sheet of Fermi surface due to its complex geometry. The calculated Fermi surfaces are displayed in (g) for the localized case, and (h) for the itinerant case. 
	}
	\label{Fig_Fermi_surface}
\end{figure*}

Fast Fourier transform (FFT) analyses of the oscillations for the two field directions are shown in Figs.~\ref{Fig_QO_and_FFT}(a) and \ref{Fig_QO_and_FFT}(c). The spectra reveal four principal frequencies, labeled $\alpha, \alpha ', \beta $ and $\beta '$. For magnetic fields applied in the $ab$-plane, both the oscillatory patterns and the corresponding FFT frequencies exhibit a slight yet systematic temperature dependence. Similar behavior has been reported in various topological materials~\cite{Lifshitz_npj2019,Lifshitz_npj2023,Lifshitz_sciadv2021} and is commonly attributed to a temperature-dependent shift of the chemical potential, leading to small changes in the extremal Fermi-surface areas. The temperature dependence of the oscillation amplitude was analyzed using the thermal damping factor  $R_{\rm{T}}$ from the Lifshitz-Kosevich (LK) formula~\cite{QO}: $R_{\rm{T}} = (\rm{X}) / \sinh(\rm{X})$, where $\rm{X}$ $= 14.69 pm^* T / B$, $p$ denotes the harmonic index and $1/B$ is the average inverse field of the FFT window. The extracted cyclotron masses $m^*$ range from 0.37(2)~$m_e$ to 0.70(1)~$m_e$, as summarized in Table~\ref{tab1}.

Subsequent high-field MR and TDO measurements
with the magnetic field along different orientations were
performed on Ce$_2$Zn$_6$Ge$_3$. Figure~\ref{Fig_high_field_QO} shows the quantum oscillations and the corresponding FFT analyses for different field orientations of out-of-plane ($\theta$) and in-plane ($\phi$) angles. The angular dependence of the oscillation frequencies is summarized in Figs.~\ref{Fig_Fermi_surface}(a) and \ref{Fig_Fermi_surface}(b) using polar coordinates. Overall, the high-field results are consistent with the low-field data. The harmonic oscillations observed in the low-field TDO measurements are not resolved in the high-field measurements, which is likely attributed to the increased noise level and lower signal resolutions. The frequencies labeled $\beta$ and $\beta'$ ($\alpha$ and $\alpha_1'$) are the Fermi surfaces split by ferromagnetism and spin-orbit coupling (SOC); $\alpha_1'$ and $\alpha_2'$ are identified as the minimum and maximum extremal cross-sections of the same Fermi surface pocket, which are supported by the calculations presented in the following section.

\subsection{Fermi surfaces with localized 4$f$ electrons}

Figures~\ref{Fig_Fermi_surface}(a) and \ref{Fig_Fermi_surface}(b) plot the experimentally derived  quantum oscillation frequencies as a function of the in-plane field angles ($\phi$) and the out-of-plane field angle ($\theta$) for Ce$_2$Zn$_6$Ge$_3$, respectively. The experimental data has been expanded across the full
angular range according to the lattice symmetry. We compare these results with the DFT calculations for the following two cases: fully localized and fully itinerant Ce-$4f$ electrons. In the localized case, the $4f$ electrons are treated as part of the core, while in the itinerant case they are included as valence electrons. In both  scenarios, the calculations were performed in the paramagnetic state and include SOC, and the SOC leads to splitting the Fermi surfaces.

As shown in Fig.~\ref{Fig_Fermi_surface}, one can see that the experimentally derived frequencies are more compatible with the scenario of assuming localized $f$-electrons in the calculations. Two low-frequency branches correspond to the minimum cross-sections of the $\alpha$ and $\alpha'$pockets, respectively, which remain nearly independent of field direction. For in-plane fields, although calculations predict hexagram-shaped pockets, the angular variations are weak; consequently, the experimental observations appear nearly isotropic. The calculated splitting between $\alpha$ and $\alpha'$ ($\sim$40 T) is notably smaller than the experimental value ($\sim$200 T). This difference likely stems from the paramagnetic state used in the calculations, which underestimates the exchange splitting present in the actual ferromagnetic phase of Ce$_2$Zn$_6$Ge$_3$. Furthermore, the calculations predict additional frequencies ($\sim$700 T) corresponding to the maximum cross-sections of the $\alpha$ and $\alpha'$ pockets, which are partially observed in the experimental results at $\phi$ $\approx$ 0$^\circ$ (and other symmetry-equivalent positions). The absence of certain frequencies may be attributed to their overlap with the $\beta$ pockets, rendering them indistinguishable. Regarding the out-of-plane results, the calculations predict that the Fermi surfaces of $\alpha$ and $\alpha'$ should be absent for fields near the $a$-axis, whereas the experimental data suggest the presence of closed pockets. On the other hand, calculations assuming localized 4$f$ electrons reveal drum-shaped Fermi surfaces for the $\beta$ and $\beta'$ pockets, which are reasonably consistent with experimental observations. However, the measured volumes of these pockets are slightly smaller than the calculated values, likely due to the influence of electronic correlations on the Fermi surfaces, despite the weak hybridization.

In contrast, the angular dependence of the calculated frequencies for the itinerant $4f$ case differs significantly from the experimental observations. First, the itinerant scenario predicts six branches of Fermi surface whereas the experiments observe only four branches. Second, the calculated frequencies and the corresponding Fermi surface are much more complicated than the experimental results as shown in Figs. \ref{Fig_Fermi_surface}(a) and \ref{Fig_Fermi_surface}(b). The above findings suggest that the ferromagnetism in Ce$_2$Zn$_6$Ge$_3$ is of a localized nature and the Kondo hybridization is weak between the Ce-$4f$ and the itinerant electrons, which is consistent with the small Sommerfeld coefficient extracted from the specific heat data.

Localized ferromagnetism and magnetic anisotropy are believed to play essential roles in the emergence of FM quantum critical points (QCPs) \cite{CeRh6Ge4_Shen2020,FMQCP_theory_PhysRevLett.120.157206,FMQCP_PhysRevB.104.L140411,FS_compare_scibullet2021}. Our results indicate that Ce$_2$Zn$_6$Ge$_3$ possesses both characteristics, suggesting that the system might be a candidate for studying ferromagnetic quantum criticality, analogous to CeRh$_6$Ge$_4$~\cite{CeRh6Ge4_Shen2020}. However, several key distinctions also exist between these two compounds. For instance, Ce$_2$Zn$_6$Ge$_3$ exhibits an easy magnetization axis along the $c$-direction, whereas in CeRh$_6$Ge$_4$ the easy axis is in the $ab$-plane, which has been proposed as a critical factor to form entangled triplet valence bond states and therefore generate strong zero-point fluctuations in a ferromagnet, giving rise to an FM QCP and the associated strange metallicity \cite{CeRh6Ge4_Shen2020}. Additionally, Ce$_2$Zn$_6$Ge$_3$ exhibits highly localized Ce-$4f$ electrons, while CeRh$_6$Ge$_4$ is a heavy fermion compound with significant hybridization between the Ce-$4f$ electrons and conduction electrons. Thus, further pressure studies exploring a possible FM QCP in Ce$_2$Zn$_6$Ge$_3$ may provide useful insights into the physics of FM quantum criticality.

\subsection{Topological properties}
\begin{figure}[H]	\includegraphics[width=1\columnwidth]{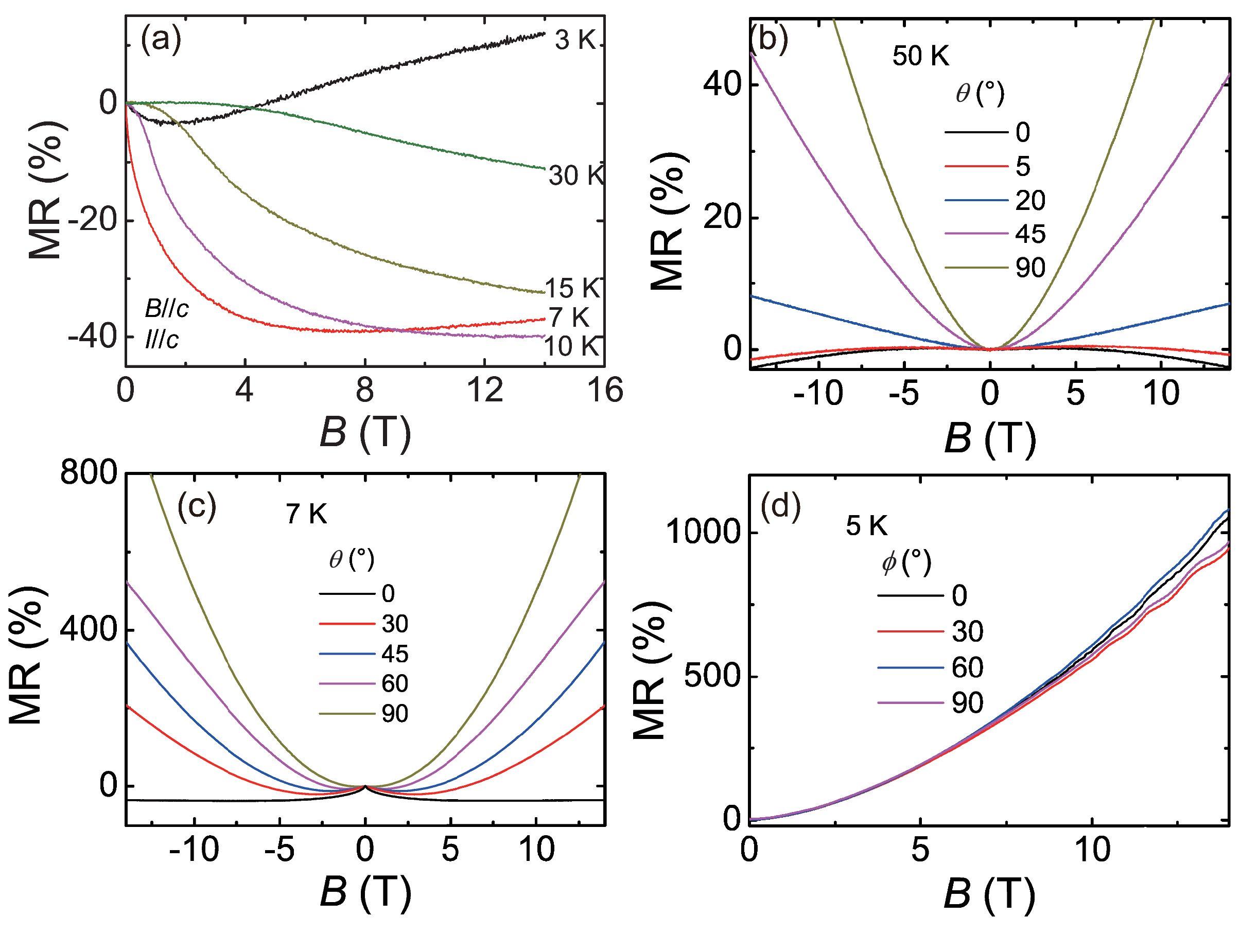} 
	\protect\caption{
		The isothermal magnetoresistance of Ce$_2$Zn$_6$Ge$_3$ at different field tilting angles; here the current is always applied along the $c$-axis. (a) The longitudinal MR ($B$$\parallel$$I$$\parallel$$c$) at selected temperatures. (b) and (c) The MR at 50 K and 7 K respectively, measured at various out-of-plane angles $\theta$, where $\theta$ is the angle between the field direction and the $c$-axis. (d) The MR at 5 K with applied magnetic field at various in-plane angles $\phi$, where $\phi$ is the angle between the field and the $a$-axis.
	}
	\label{Fig_MR_rotate}
\end{figure}

Figure~\ref{Fig_MR_rotate}(a) illustrates the MR measurements at various temperatures with both the current and magnetic field  applied along the $c$-axis ($B$$\parallel$$I$$\parallel$$c$), revealing a negative longitudinal magnetoresistance (LMR). In contrast, a large positive MR is observed when the current is parallel to the $c$-axis and the field is within the $ab$-plane, as shown in Fig.~\ref{Fig_QO_raw_data}(a). Additionally, the MR measurements at different field angles tilted from the $c$-axis are performed at temperatures both above [Fig.~\ref{Fig_MR_rotate}(b)] and below [Fig.~\ref{Fig_MR_rotate}(c)] the ferromagnetic transition. At 50~K, a negative LMR is observed up to at least 14~T. As the field is applied slightly away from the $c$-axis, a positive MR develops, which becomes more pronounced as the field approaches the $ab$-plane. Upon further decreasing temperature, the transverse MR ($B$$\perp$$I$) is significantly enhanced, while the negative LMR changes its behavior at temperatures below $T_{\rm C}$. As shown in Fig. \ref{Fig_MR_rotate}(a), the LMR shows a minimum and then eventually becomes positive with increasing magnetic field at temperatures below $T_{\rm C}$, which is likely attributed to the magnetic scattering in the ordered state. Furthermore, Fig.~\ref{Fig_MR_rotate}(d) displays the MR with fields applied within the $ab$-plane, showing that the planar MR is nearly angle independent. These results are consistent with the presence of topologically nontrivial electronic states in Ce$_2$Zn$_6$Ge$_3$. However, we note that negative LMR alone does not constitute definitive evidence for a chiral anomaly; additional experiments are needed to verify the topological state.

\begin{figure}[H]
	\includegraphics[width=0.65\columnwidth]{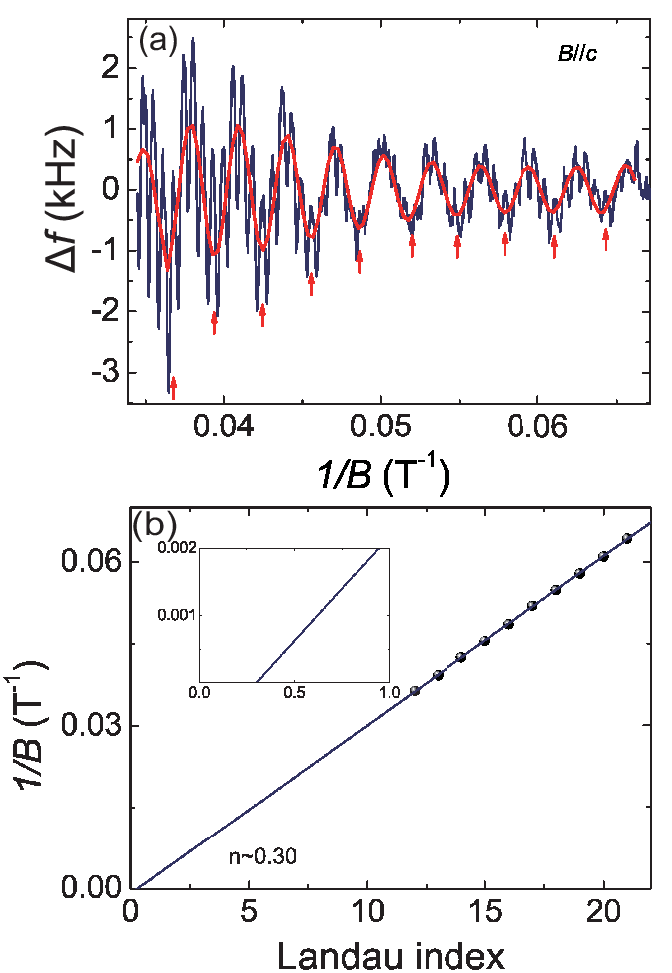} 
	\centering\protect\caption{
		(a)	The high-field TDO quantum oscillations ($B$$\parallel$$c$) at 60 mK, plotted against 1/$B$. The red line shows a good fit of the quantum oscillations ($\alpha$ branch) by using the LK formula. (b) The dependence of the Landau index on 1/$B$, obtained from the positions of the minimum of the oscillations. The solid line shows a linear fit and yields a residual Landau index of n$_0$ $\sim$ 0.30. The inset shows an enlarged view of the intercept.
	}
	\label{Fig_landau_index}
\end{figure}

High-field quantum oscillations are further utilized to characterize the possible topological state of Ce$_2$Zn$_6$Ge$_3$. In Fig.~\ref{Fig_landau_index}(a), the TDO data at 60~mK are plotted as a function

\noindent
of $1/B$ ($B$$\parallel$$c$), which are fitted by using the LK formula~\cite{QO, QO_RevModPhys.82.1959}:

\begin{equation}
	\Delta\rho/\rho_0 = \sqrt{B} R_{\rm{T}} R_{\rm{D}} R_{\rm{S}} \cos\left[2\pi \left(\frac{F}{B} - \phi\right)\right]. \nonumber
\end{equation}

Here, $R_{\rm{T}} = (14.69 m^* T / B) / \sinh(14.69 m^* T / B)$, $R_{\rm{D}} = \exp(-14.69 m^* T_{\rm{D}} / B)$, $R_{\rm{S}} = \cos(\pi g m^* / 2m_e)$ refer to the thermal, Dingle and spin damping factors, respectively. The variables $m^*$ and $m_e$ denote the cyclotron mass and the free electron mass, respectively; $T_{\rm{D}}$ is the Dingle temperature, and $g$ is the $g$-factor. The total phase is given by $\phi = \phi_{\rm{B}}/2\pi + \phi_{\rm{D}}$, where $\phi_{\rm{B}}$ is the Berry phase and $\phi_{\rm{D}}$ is an additional phase shift. For the cross-section of a three-dimensional (3D) Fermi surface, $\phi_{\rm{D}}$ = $-1/8$ (corresponding to a $-\pi/4$ phase shift in the cosine)~\cite{Landua_fan_berry_phase_PRL1999}. As shown in Fig. \ref{Fig_landau_index}(a), the LK formula can well describe the wave packet of the low-frequency oscillation ($\alpha$ branch), giving a Berry phase of 0.83(6)$\pi$.

A Landau fan diagram~\cite{Landua_fan_berry_phase_PRL2018,Landau_fan_SCPMA2026} is also employed to extract the oscillation phase factor, where the oscillatory valleys are assigned to the integral Landau indices. Linear extrapolation of the Landau indices yields an intercept of n$_0 = 0.30$, giving a Berry phase $\phi_{\rm B} = 2\pi (n_0 - \phi_{\rm{D}}$) = 0.85$\pi$. Therefore, the analyses based on the fits of the LK formula and the Landau fan diagram yield a Berry phase close to $\pi$. These results are consistently indicative of a nontrivial Berry phase and are thus compatible with the existence of topologically nontrivial electronic states in Ce$_2$Zn$_6$Ge$_3$.

\begin{figure}[H]
	\includegraphics[width=0.7\columnwidth]{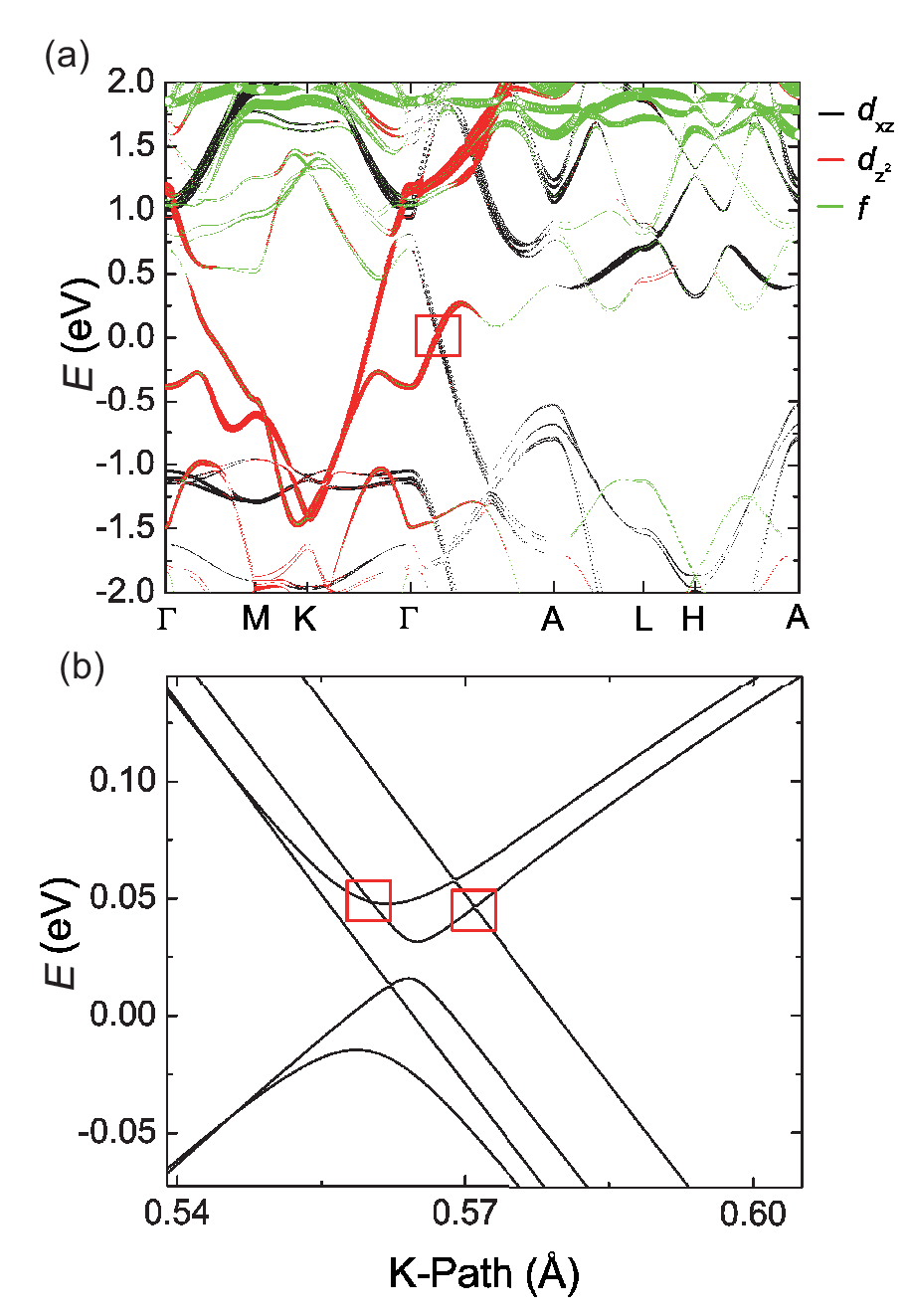} 
	\centering\protect\caption{
		(a) Band structure of Ce$_2$Zn$_6$Ge$_3$ from the contribution of Ce-$d_{xz}$ (black), $d_{z^2}$ (red) and $f$ (green) under the FM order configuration and with the inclusion of SOC and Coulomb interaction $U$. (b) Magnified view of the band structure near the Fermi level, where two pairs of Weyl points are indicated by red squares.
	}
	\label{Fig_calculation}
\end{figure}

To further determine the nature of the topological state, the band structure of Ce$_2$Zn$_6$Ge$_3$ was calculated using the DFT calculations within the GGA$+U$ method. Here an on-site Coulomb interaction $U = 6$~eV and an exchange interaction $J = 0.7$~eV were applied to the Ce-4$f$ orbitals. The Brillouin zone was sampled with a $6 \times 6 \times 10$ $k$-mesh for self-consistent calculations and a $12 \times 12 \times 20$ $k$-mesh for the calculations of the density of states (DOS). In the calculations, SOC and FM order aligned with the $c$-axis are included. It should be noted that in our calculations, various magnetic configurations are considered and the energetically favorable ground state is the one with moments aligned along the $c$-axis. On the other hand, the principal features of the bands remain largely robust regardless of the specific magnetic configuration. As illustrated in Fig. \ref{Fig_calculation}, the band structure exhibits linear dispersion, with band crossings forming two pairs of Weyl points in close proximity to the Fermi level. The GGA$+U$ calculations, where the Ce-4$f$ electrons are located far away from the Fermi level, demonstrate a high degree of similarity to the calculations for fully localized 4$f$ electrons presented in Section 3.2. The Weyl fermions are primarily contributed by the Ce-$d$ orbits, specifically the $d_{z^2}$, $d_{xz}$, and $d_{yz}$ orbitals for fully localized 4$f$ electrons which place the $f$-electron contributions far from the Fermi level. This finding underscores the significance of the Ce-based honeycomb lattice in hosting topologically nontrivial band structures.

\section{Conclusions}\label{sec:4}
In summary, we have investigated the Fermi surface and topological properties of the Ce-based honeycomb ferromagnet Ce$_2$Zn$_6$Ge$_3$ by measuring the transport properties and quantum oscillations up to 30 T in combination with the DFT calculations. The experimentally derived quantum oscillation frequencies are compatible with the band structure calculations assuming localized Ce-4$f$ electrons, suggesting the existence of localized ferromagnetism with strong anisotropy in Ce$_2$Zn$_6$Ge$_3$. Furthermore, the observations of the negative longitudinal MR, a Berry phase close to $\pi$ in quantum oscillations and the existence of two pairs of Weyl points near the Fermi level in the calculated band structure have clearly established Ce$_2$Zn$_6$Ge$_3$ as a topological material with nontrivial properties. Therefore, Ce$_2$Zn$_6$Ge$_3$ would provide an ideal platform for studying topology, magnetism, quantum criticality and their interplay. Further studies, such as tuning the ground state via pressure or doping, along with various spectroscopic measurements, are warranted to elucidate these properties. 

\Acknowledgements{This work was supported by the National Key R\&D Program of China (Grant No. 2022YFA1402200 and No. 2023YFA1406303), the National Natural Science Foundation of China (Grant No. 12550401, No. W2511006, No. 12674175, No. 12274364 and No. 12494592) and the New Cornerstone Science Foundation (Grant No. NCI202509). A portion of this work was carried out at the Synergetic Extreme Condition User Facility (SECUF, https://cstr.cn/31123.02.SECUF).}

\InterestConflict{The authors declare that they have no conflict of interest.}

\end{multicols}
\end{document}